\documentclass[aps, prd
, twocolumn
, nofootinbib
, superscriptaddress
, preprintnumbers
]{revtex4-2}

\usepackage{graphicx}
\usepackage{xcolor}
\usepackage{mathrsfs,mathtools}
\usepackage{physics,amssymb}
\usepackage{siunitx}
\usepackage{bm}
\usepackage{soul}
\usepackage{cancel}
\usepackage{url}
\usepackage{longtable}
\usepackage{xspace}
\usepackage{acronym}
\usepackage[colorlinks=true
,urlcolor=DARKBLUE
,anchorcolor=DARKBLUE
,citecolor=DARKBLUE
,filecolor=DARKBLUE
,linkcolor=DARKBLUE
,menucolor=DARKBLUE
,linktocpage=true
,pdfproducer=medialab
,pdfa=true
]{hyperref}

\newcommand{\ee}{\mathrm{e}}
\newcommand{\Mpl}{M_\mathrm{Pl}}

\newcommand{\gO}{\mathrm{O}}
\newcommand{\water}{\text{water}}
\newcommand{\FP}{\mathrm{FP}}
\newcommand{\FPT}{\mathrm{FPT}}

\newcommand{\eigenL}{\lambda}

\newcommand{\uc}{\mathrm{c}}

\newcommand{\bfk}{\mathbf{k}}

\newcommand{\calL}{\mathcal{L}}

\newcommand{\calN}{\mathcal{N}}

\newcommand{\calP}{\mathcal{P}}

\newcommand{\ur}{\mathrm{r}}

\newcommand{\us}{\mathrm{s}}

\newcommand{\beae}[1]{\begin{equation}\begin{aligned} #1 \end{aligned}\end{equation}}
\newcommand{\bege}[1]{\begin{equation}\begin{gathered} #1 \end{gathered}\end{equation}}
\newcommand{\bae}[1]{\begin{align} #1 \end{align}}

\newcommand{\bme}[1]{\begin{multline} #1 \end{multline}}

\definecolor{MONZA}{HTML}{CF000F}
\definecolor{DARKBLUE}{HTML}{00008b}
\definecolor{DARKMAGENTA}{HTML}{8b008b}

\begin{document}
\title{Stochastic-tail of the curvature perturbation in hybrid inflation}
\date{\today}

\author{Tomoaki Murata}
\email{tomoaki-m@metro-cit.ac.jp}
\affiliation{Department of General Education, Tokyo Metropolitan College of Industrial Technology, 
Shinagawa, Tokyo 140-0011, Japan}
\affiliation{Department of Physics, Rikkyo University, Toshima, Tokyo 171-8501, Japan}

\author{Yuichiro Tada}
\email{yuichiro.tada@rikkyo.ac.jp}
\affiliation{Department of Physics, Rikkyo University, Toshima, Tokyo 171-8501, Japan}

\begin{abstract}
The exponential-tail behaviours of the \ac{PDF} of the primordial curvature perturbation are confirmed in the mild-waterfall variants of hybrid inflation with the use of the stochastic formalism of inflation.
On top of these tails, effective upper bounds on the curvature perturbation are also observed, corresponding to the exact hilltop trajectory during the waterfall phase.
We find that in the model where the leading and higher-order terms in the expansion of the inflaton potential around the critical point are fine-tuned to balance, this upper bound can be significantly reduced, even smaller than the \ac{PBH} threshold, as a novel perturbation-reduction mechanism than the one proposed in Ref.~\cite{Tada:2023pue}.
It makes \ac{PBH} formation much difficult compared to the Gaussian or exponential-tail approximation. 
We also introduce Johnson's $S_U$-distribution as a useful fitting function for the \ac{PDF}, which reveals a nonlinear mapping between the Gaussian field and the curvature perturbation. 
\end{abstract}

\preprint{RUP-25-17}
\maketitle

\acrodef{PBH}{primordial black hole}
\acrodef{MD1}{mass-dimension $1$}
\acrodef{CMB}{cosmic microwave background}
\acrodef{PDF}{probability density function}
\acrodef{EoM}{equation of motion}
\acrodef{SIGW}{scalar-induced gravitational wave}

\acresetall
\section{Introduction}

Cosmic inflation has been studied from various aspects so far as a leading paradigm of the early universe.
In particular, the properties of the primordial curvature perturbation $\zeta$ generated during inflation are recorded as cosmological/astrophysical imprints in many ways so that they can be direct observables.
The \emph{exponential-tail} behaviour is one of the most interesting features amongst such properties of the curvature perturbation~\cite{Pattison:2017mbe,Ezquiaga:2019ftu,Figueroa:2020jkf,Pattison:2021oen,Cai:2018dkf,Atal:2019cdz,Atal:2019erb,Pi:2021dft,Pi:2022ysn,Wang:2024xdl,Inui:2024sce,Inui:2024fgk}.
It suggests that the large value probability of the curvature perturbation decays only exponentially $\sim\ee^{-\eigenL\zeta}$ in a wide class of inflation models, instead of the ordinary Gaussian distribution $\sim\ee^{-\zeta^2/(2\sigma^2)}$.
It can significantly affect the formation of rare objects such as \acp{PBH}~\cite{Biagetti:2021eep,Kitajima:2021fpq,Tada:2021zzj}, massive galaxy clusters~\cite{Ezquiaga:2022qpw}, and so on.

Though the exponential tail has been studied well in single-field inflation models, it is also expected to be realised in multi-field models.
In this paper, we focus on \emph{hybrid inflation}~\cite{Linde:1993cn} as a representative multi-field model.
Hybrid inflation is the ``hybrid'' of chaotic inflation and hilltop inflation: the inflaton first rolls down its potential towards the origin, called the \emph{valley} phase, and then the other waterfall fields are triggered to roll down their potential like hilltop inflation to end inflation, called the \emph{waterfall} phase.
The dynamics of the waterfall phase are strongly affected by the fluctuations of the waterfall fields at the onset of the phase (critical point), indicating the non-perturbative nature of the system.
The resultant curvature perturbation is hence expected to be sizable, so the \ac{PBH} formation in this model is an exciting topic, particularly in the \emph{mild-waterfall} variants where the waterfall phase lasts for a few or tens e-folds to make the formed \acp{PBH} massive enough (see, e.g., Refs.~\cite{Garcia-Bellido:1996mdl,Lyth:2010zq,Bugaev:2011qt,Bugaev:2011wy,Lyth:2012yp,Guth:2012we,Halpern:2014mca,Clesse:2015wea,Kawasaki:2015ppx,Tada:2023pue,Tada:2023fvd,Tada:2024ckk}).
One interesting feature of the mild-waterfall models is that the model parameters partially degenerate, and there is a linear relation between the expected duration of the waterfall phase and the amplitude of the power spectrum of the curvature perturbation~\cite{Clesse:2015wea,Kawasaki:2015ppx,Tada:2023pue}.
Calculating the perturbation up to the kurtosis, Ref.~\cite{Kawasaki:2015ppx} concluded that massive enough \acp{PBH} are inevitably overproduced in the natural setup of the mild-waterfall hybrid inflation with a few waterfall fields.
To resolve the degeneracy, one can increase the number of waterfall fields~\cite{Halpern:2014mca}, or introduce a fine-tuned balance between the leading and higher-order terms in the expansion of the inflaton potential around the critical point.
In Refs.~\cite{Tada:2023pue,Tada:2023fvd,Tada:2024ckk}, Tada, one of the authors, has investigated these possibilities at the level of the power spectrum of the curvature perturbation with Yamada.
In particular, Ref.~\cite{Tada:2023pue} (TY23a hereafter) studied the reduction of the power spectrum in the balanced model by generalising the semi-perturbative approach of Refs.~\cite{Clesse:2015wea,Kawasaki:2015ppx}.

In this paper, we investigate the \ac{PDF} of the curvature perturbation in the mild-waterfall hybrid inflation beyond the power spectrum with the use of the \emph{stochastic formalism} of inflation~\cite{Starobinsky:1982ee,Starobinsky:1986fx,Starobinsky:1994bd} (see also references in the recent review~\cite{Cruces:2022imf}).
In this formalism, the curvature perturbation is identified with the fluctuation in the first passage time $\calN$ in terms of the e-folding number from a certain initial field value to the end surface of inflation along the stochastic dynamics of the inflaton/waterfall fields, known as the \emph{stochastic-$\delta\calN$} approach~\cite{Fujita:2013cna,Fujita:2014tja,Vennin:2015hra,Ando:2020fjm,Animali:2024jiz}.
Not only do we confirm the exponential-tail behaviour of the \ac{PDF}, but we also find an effective upper bound on the curvature perturbation beyond which the probability of large perturbations is strongly suppressed.
Such an upper bound is significant in the balanced model, and consequently, the perturbation variance is much reduced compared to TY23a.

We also introduce Johnson's $S_U$-distribution~\cite{9d62cdcd-54fa-36dd-8d61-6ad5b637bd6e,10.1093/biomet/36.3-4.297} as a useful fitting function both for the peak and tail behaviours of the \ac{PDF}.
It reveals a nonlinear mapping between the Gaussian field and the curvature perturbation, which will be helpful for the application of the peak theory~\cite{Bardeen:1985tr,Yoo:2018kvb,Yoo:2020dkz} to estimate the \ac{PBH} mass function (see, e.g., Refs.~\cite{Yoo:2019pma,Kitajima:2021fpq,Inui:2024fgk} for applications of nonlinear mappings to the peak theory).

The paper is organised as follows.
In Sec.~\ref{sec: stoc-dN}, we review the stochastic-$\delta\calN$ approach and the exponential-tail feature of the curvature perturbation first, and then introduce Johnson's $S_U$-distribution.
In Sec.~\ref{sec: hybrid}, we show our results on the \ac{PDF} in several setups of the mild-waterfall hybrid inflation.
They are compared with TY23a in the last subsection.
Sec.~\ref{sec: D and C} is devoted to discussion and conclusions.
Throughout the paper, we adopt the natural unit where $c=\hbar=1$.

\section{\boldmath Stochastic-$\delta\calN$ approach to the probability distribution of the curvature perturbation}\label{sec: stoc-dN}

\subsection{\boldmath Stochastic-$\delta\calN$ approach and the exponential tail}\label{eq: exp-tail}

The primordial perturbation in cosmic inflation is often calculated using the quantum theory of perturbed fields on a spatially homogeneous, classical background.
However, it is not necessarily justified to suppose a spatially homogeneous background because any two spatial points further than the Hubble scale are causally decoupled.
In fact, field values at separated two points can have significantly different values in cases where the perturbations non-negligibly backreact on the background, which can happen, e.g., in models of \ac{PBH} formation.
The so-called \emph{stochastic formalism}~\cite{Starobinsky:1982ee,Starobinsky:1986fx,Starobinsky:1994bd} (see also references in the recent review~\cite{Cruces:2022imf}) is helpful in these cases.
This is an effective theory of fields coarse-grained on a certain superHubble scale.
With the use of the gradient expansion and the assumption that the quantumness of the superHubble coarse-grained fields is negligible, it is justified to view the dynamics of the coarse-grained fields as a locally flat universe with classical, stochastic noise caused by the horizon crossing of the subHubble fields.
That is, the field equations for the coarse-grained system are given at each spatial point by\footnote{For a spatial correlation, see the discussions in Refs.~\cite{Fujita:2013cna,Fujita:2014tja,Vennin:2015hra,Ando:2020fjm,Animali:2024jiz}, a numerical lattice implementation~\cite{Mizuguchi:2024kbl}, a box assignment simulation~\cite{Animali:2025pyf}, etc. Also, note that $\pi_i$ is not the momentum conjugate with respect to the e-folds $N$, but is rather related to the cosmic time derivative, or the first equation in \eqref{eq: EoM} should be understood as its definition. This definition simplifies the Friedmann equation (the last equation). See, e.g., Ref.~\cite{Pinol:2020cdp} for more details.}
\bege{\label{eq: EoM}
    \dv{\phi_i}{N}=\frac{\pi_i}{H}+\xi_{\phi_i} \qc
    \dv{\pi_i}{N}=-3\pi_i-\frac{V_i}{H}+\xi_{\pi_i}, \\
    3\Mpl^2H^2=\frac{\pi_i^2}{2}+V.
}
The subscript $i=1,2,\dots$ distinguishes the inflaton fields if there are multiple ones.\footnote{Though we suppose that the inflatons span a flat target manifold, the stochastic formalism can be generalised to a curved target field (see Ref.~\cite{Pinol:2020cdp}).}
$V_i=\partial_{\phi_i}V$ is the field derivative of the potential $V$, $H$ is the Hubble parameter, $N$ is the e-folding number as the time variable,\footnote{See, e.g., Refs.~\cite{Finelli:2008zg,Finelli:2010sh,Pattison:2019hef} for discussions about e-folding number as the time variable in the stochastic formalism.} and $\Mpl\simeq\SI{2.4e18}{GeV}$ is the reduced Planck mass.
$\xi_\phi$ and $\xi_\pi$ represent the stochastic white noises, whose statistics are characterised by
\bme{
    \expval{\xi_X}=0 \qc
    \expval{\xi_X(N)\xi_Y(N')}=\calP_{XY}\delta(N-N'), \\
    \text{for} \quad X,Y=\phi_i,\pi_i.
}
Here, the (log-bin) power spectrum $\calP_{XY}$ defined by
\bae{\!\!\!
    \expval{X_\bfk(N)Y_{\bfk'}(N)}=(2\pi)^3\delta^{(3)}(\bfk+\bfk')\frac{2\pi^2}{k^3}\calP_{XY}(k,N),
}
is evaluated at the coarse-graining scale.
In the rest of the paper, we adopt the following slow-roll approximation under the flat target manifold assumption: 
\bae{\label{eq: SR noise}
    \calP_{\phi_i\phi_j}\approx\pqty{\frac{H}{2\pi}}^2\delta_{ij} \qc
    (\text{otherwise})\approx0.
} 
There, the stochastic noises in the \ac{EoM}~\eqref{eq: EoM} can be replaced by
\bae{
    \xi_{\phi_i}=\frac{H}{2\pi}\xi_i \qc \xi_{\pi_i}=0,
}
with the normalised noise
\bae{
    \expval{\xi_i}=0 \qc
    \expval{\xi_i(N)\xi_j(N')}=\delta_{ij}\delta(N-N').
}

The stochastic formalism provides a probability distribution of the inflatons' fluctuations, but it should be converted to that of the curvature perturbation $\zeta$ conserved after inflation, as the inflatons themselves have disappeared.
It should be noted that the duration of inflation at each spatial point is no longer deterministic but stochastically fluctuates due to the noise.
According to the $\delta N$ formalism~\cite{Starobinsky:1985ibc,Lyth:2004gb}, such a fluctuation in terms of the e-folds is nothing but the curvature perturbation.
This coincidence has inspired the \emph{stochastic-$\delta\calN$} approach~\cite{Fujita:2013cna,Fujita:2014tja,Vennin:2015hra,Ando:2020fjm,Animali:2024jiz} (see also Refs.~\cite{Kawasaki:2015ppx,Assadullahi:2016gkk,Vennin:2016wnk,Pattison:2017mbe,Ezquiaga:2018gbw,Noorbala:2018zlv,Firouzjahi:2018vet,Noorbala:2019kdd,Kitajima:2019ibn,Prokopec:2019srf,Ezquiaga:2019ftu,Firouzjahi:2020jrj,De:2020hdo,Figueroa:2020jkf,Pattison:2021oen,Figueroa:2021zah,Tada:2021zzj,Ezquiaga:2022qpw,Ahmadi:2022lsm,Nassiri-Rad:2022azj,Animali:2022otk,Tomberg:2022mkt,Gow:2022jfb,Rigopoulos:2022gso,Briaud:2023eae,Asadi:2023flu,Tomberg:2023kli,Tada:2023fvd,Tokeshi:2023swe,Raatikainen:2023bzk,Miyamoto:2024hin,Tokeshi:2024kuv,Kuroda:2025coa,Takahashi:2025hqt} for its application).
In this approach, the curvature perturbation is calculated as a fluctuation in the first passage time $\calN$ (i.e., $\zeta=\delta\calN$) from a certain initial field value to the end surface of inflation.\footnote{Conventionally, the ordinary italic $N$ is used for the deterministic, forward e-folding number as a time variable, while the calligraphic $\calN$ is used for the stochastic first passage time.}
The problem hence reduces to the probability distribution of the first passage time of the stochastic process~\eqref{eq: EoM}.

Under the Markovian approximation represented by the slow-roll approximation of the noise amplitude~\eqref{eq: SR noise} (i.e., the noise amplitude is determined only by the current values of the inflatons and momenta, independently of the past history), the \ac{PDF} of the inflatons and momenta is known to follow the Fokker--Planck equation\footnote{To derive the Fokker--Planck equation, the noise should be interpreted in It\^o's way~\cite{Tokuda:2017fdh,Tokuda:2018eqs,Pinol:2020cdp}.}
\bae{
    \partial_NP(\bm{\phi},\bm{\pi}\mid N)=\calL_\FP P(\bm{\phi},\bm{\pi}\mid N),
}
where $\calL_\FP$ a second-order elliptic partial differential operator in $\bm{\phi}=(\phi_1,\phi_2,\dots)$ and $\bm{\pi}=(\pi_1,\pi_2,\dots)$.
Ref.~\cite{Vennin:2015hra} has shown that if the \ac{PDF} of the inflatons follows the Fokker--Planck equation, the counterpart of the first passage time $\calN$ from initial values $\bm{\phi}$ and $\bm{\pi}$ follows its adjoint equation:
\bae{
    \partial_\calN P_\FPT(\calN\mid\bm{\phi},\bm{\pi})=\calL_\FP^\dagger P_\FPT(\calN\mid\bm{\phi},\bm{\pi}),
}
with the adjoint Fokker--Planck operator $\calL_\FP^\dagger$ defined by
\bme{
    \int_\Omega\dd{\bm{\phi}}\dd{\bm{\pi}}f(\bm{\phi},\bm{\pi})\calL_\FP g(\bm{\phi},\bm{\pi}) \\
    =\int_\Omega\dd{\bm{\phi}}\dd{\bm{\pi}}g(\bm{\phi},\bm{\pi})\calL_\FP^\dagger f(\bm{\phi},\bm{\pi}),
}
for arbitrary functions $f$ and $g$ on the relevant domain $\Omega$ in the phase space.
If the (stochastic) inflation domain $\Omega$ is (effectively) compact, the adjoint Fokker--Planck operator has discrete eigenvalues $-\eigenL_n$ (a minus sign is introduced for later convenience),\footnote{In the slow-roll limit, the adjoint Fokker--Planck operator reduces to
\bae{
    \frac{1}{\Mpl^2}\calL_\FP^\dagger=-\sum_i\frac{v_i}{v}\partial_{\phi_i}+v\sum_i\partial_{\phi_i}^2 \qc
    v(\bm{\phi})=\frac{V(\bm{\phi})}{24\pi^2\Mpl^4},
}
where the momenta are irrelevant. In this case, one can define a Hermitian operator $\widetilde{\calL_\FP}$ with the same eigenvalues as $\calL_\FP^\dagger$ by~\cite{Ezquiaga:2019ftu,Miyamoto:2024hin}
\bae{
    \widetilde{\calL_\FP}=w^{1/2}(\bm{\phi})\calL_\FP^\dagger w^{-1/2}(\bm{\phi}) \qc
    w(\bm{\phi})=\frac{\ee^{1/v(\bm{\phi})}}{v(\bm{\phi})}.
}
Therefore, it can be shown that the corresponding eigenvalues are real.
}
and the \ac{PDF} can be expanded by the corresponding eigenfunctions $\psi_n(\bm{\phi},\bm{\pi})$ as
\bae{
    P_\FPT(\calN\mid\bm{\phi},\bm{\pi})=\sum_{n\geq0}\alpha_n\psi_n(\bm{\phi},\bm{\pi})\ee^{-\eigenL_n\calN}.
}
This expression indicates that the \ac{PDF} exhibits an exponential decay at large $\calN$ (and hence large curvature perturbation) with the decay constant $\eigenL_0$, the lowest eigenvalue, which can be much slower than the ordinary Gaussian distribution $\propto\ee^{-\delta\calN^2/(2\sigma^2)}$.
This non-perturbative behaviour of the tail is called the \emph{exponential tail}~\cite{Pattison:2017mbe,Ezquiaga:2019ftu,Figueroa:2020jkf,Pattison:2021oen}.\footnote{Similar exponential tails have been found also in the non-stochastic $\delta N$ approach~\cite{Cai:2018dkf,Atal:2019cdz,Atal:2019erb,Pi:2021dft,Pi:2022ysn,Wang:2024xdl,Inui:2024sce} and summarised as the \emph{logarithmic non-Gaussianity} in Ref.~\cite{Inui:2024fgk}.}
Though it does not affect the statistics of typical, small perturbations, it can significantly alter the probability of large perturbations and hence the prediction of rare objects such as \acp{PBH}~\cite{Biagetti:2021eep,Kitajima:2021fpq,Tada:2021zzj}, massive galaxy clusters~\cite{Ezquiaga:2022qpw}, etc.
Note that if the relevant phase space $\Omega$ is non-compact, the eigenvalues of the adjoint Fokker--Planck operator are continuous in general, and it is suggested that the \ac{PDF} shows even heavier tail than the exponential tail in such a case~\cite{Vennin:2024yzl}.

\subsection{\boldmath Johnson's $S_U$-distribution}\label{sec: SU}

Over several types of the tail of the \ac{PDF}, we introduce Johnson's $S_U$-distribution~\cite{9d62cdcd-54fa-36dd-8d61-6ad5b637bd6e,10.1093/biomet/36.3-4.297} as a useful fitting function.
It is defined by a nonlinear transformation of the Gaussian distribution:
\bae{\label{eq: g to dN}
    \delta\calN=\sigma\sinh\frac{g-\gamma}{\rho}+\mu,
}
where the random variable $g$ follows the Gaussian distribution with $\expval{g}=0$ and $\expval{g^2}=1$, and $\mu$, $\sigma$, $\gamma$, and $\rho$, are the fitting parameters.
They roughly control the mean, variance, skewness, and kurtosis of the distribution, respectively.
The \ac{PDF} of this distribution is given by
\bme{\label{eq: Johnson PDF}
    P(\delta\calN)=\frac{\rho}{\sqrt{2\pi\pqty{(\delta\calN-\mu)^2+\sigma^2}}} \\
    \times\exp[-\frac{1}{2}\pqty{\gamma+\rho\sinh^{-1}\frac{\delta\calN-\mu}{\sigma}}^2].
}
Around $\delta\calN\sim\mu$ or $g\sim\gamma$, the hyperbolic sine function can be approximated as
\bae{
    \frac{\delta\calN-\mu}{\sigma}\simeq\frac{g-\gamma}{\rho},
}
and hence $\delta\calN$ follows the Gaussian distribution.
On the other hand, for a large perturbation $\delta\calN\gg\mu$, the hyperbolic sine behaves as $\sinh x\simeq\ee^x/2$ and the perturbation value is exponentially enhanced, observed as a heavy tail.
In fact, the effective decay constant defined by 
\bae{
    \eigenL\coloneqq-\dv{\ln P(\delta\calN)}{\delta\calN},
}
asymptotes to
\bae{
    \eigenL\overset{\delta\calN\gg1}{\sim}\frac{2+2\gamma\rho+2\rho^2\ln\delta\calN+\rho^2\ln(4/\sigma^2)}{2\delta\calN}\to0.
}
It is heavier than the exponential tail according to the definition of Ref.~\cite{Hooshangi:2021ubn}.
Therefore, depending on the choice of the parameters, Johnson's $S_U$-distribution can smoothly interpolate between the Gaussian (light tail, $\eigenL\propto\delta\calN$) behaviour and the heavy tail ($\eigenL\to0$) via the exponential tail ($\eigenL\to\text{const.}$).
Below, we will see that practically, the physically motivated region $\delta\calN\sim1$, e.g., for \ac{PBH} formation, corresponds to the exponential-tail part.

Once the mapping~\eqref{eq: g to dN} from the Gaussian field to the curvature perturbation has been revealed, the \emph{peak theory} (see Ref.~\cite{Bardeen:1985tr} for the original work and Refs.~\cite{Yoo:2018kvb,Yoo:2020dkz} for its application to the \ac{PBH} abundance) can be applied to calculate the \ac{PBH} mass function.
The specific computations have been shown for the quadratic non-Gaussianity~\cite{Yoo:2019pma} and the logarithmic non-Gaussianity~\cite{Kitajima:2021fpq,Inui:2024fgk}.
It is, however, beyond the scope of the paper because the non-Gaussian case with a broad power spectrum practically requires too complicated procedures.
It should also be clarified whether the spatial distribution of the curvature perturbations exhibits a certain non-trivial configuration or not, because hybrid inflation is featured by topological defects~\cite{STOLAS}.
We hence leave it for future work.

\section{Hybrid inflation}\label{sec: hybrid}

\subsection{Up to the quadratic term}

Hybrid inflation~\cite{Linde:1993cn} consists of two species of scalar fields. One is the inflaton field $\phi$, which is often supposed to be a single real scalar.
The other is the waterfall fields $\bm{\psi}=\pqty{\psi_1,\psi_2,\cdots,\psi_n}$, which are in general $n$ real scalars respecting $\gO(n)$ symmetry.
Hybrid inflation is characterised by the following potential form:
\bae{
    V(\phi,\bm{\psi})=\Lambda^4\bqty{\pqty{1-\frac{\psi_\ur^2}{M^2}}^2+2\frac{\phi^2\psi_\ur^2}{\phi_\uc^2M^2}+v(\phi)},
}
where $\Lambda$, $M$, and $\phi_\uc$ are parameters \ac{MD1}, $\psi_\ur=\sqrt{\sum_i\psi_i^2}$ is the radial direction of the waterfall fields, and $v(\phi)$ represents the inflaton's part of the potential.
One finds that while $\psi_\ur$ is massive at origin $\eval{V_{\psi_\ur\psi_\ur}}_{\psi_\ur=0}>0$ for $\phi>\phi_\uc$, it becomes unstable $\eval{V_{\psi_\ur\psi_\ur}}_{\psi_\ur=0}<0$ if $\phi<\phi_\uc$.
Therefore, if $\phi$ rolls from a larger value $\phi>\phi_\uc$ down to a small value $\phi<\phi_\uc$ along its potential $v(\phi)$, the waterfall $\psi_\ur$ is first stabilised at around the origin $\psi_\ur\sim0$ for $\phi>\phi_\uc$ (called \emph{valley phase}) but then destabilised for $\phi<\phi_\uc$, rapidly rolls down to the true potential minimum $\psi_\ur=M$, and ends the slow-roll inflation (called \emph{waterfall phase}).
This is the basic picture of hybrid inflation.

Though the waterfall phase is almost instantaneous in the standard setup, it can last for a few or tens of e-folds depending on the choice of parameters~\cite{Clesse:2010iz,Kodama:2011vs,Mulryne:2011ni,Clesse:2012dw}. We consider the \emph{mild case}, where the waterfall phase lasts for more than a few e-folds but fewer than $60$ e-folds.
In this case, the comoving Hubble size at the critical point $\phi=\phi_\uc$ lies in the observable universe, and the curvature perturbation on that scale is expected to be significantly enhanced.
The \ac{PBH} formation in this context has been extensively investigated (see, e.g., Refs.~\cite{Garcia-Bellido:1996mdl,Lyth:2010zq,Bugaev:2011qt,Bugaev:2011wy,Lyth:2012yp,Guth:2012we,Halpern:2014mca,Clesse:2015wea,Kawasaki:2015ppx,Tada:2023pue,Tada:2023fvd,Tada:2024ckk}).
There, the inflaton potential $v(\phi)$ is series-expanded around $\phi_\uc$ up to the quadratic order as
\bae{\label{eq: v quadratic}
    v(\phi)=\frac{\phi-\phi_\uc}{\mu_1}-\frac{(\phi-\phi_\uc)^2}{\mu_2^2},
}
with two \ac{MD1} parameters $\mu_1$ and $\mu_2$.

Among the four model parameters $\Lambda$, $M$, $\mu_1$, and $\mu_2$, two degrees of freedom are fixed by the \ac{CMB} anisotropy.
The primordial perturbations on the \ac{CMB} scale ($\sim\SI{0.05}{Mpc^{-1}}$) correspond to the valley phase $\phi>\phi_\uc$ where the waterfall $\psi_\ur\sim0$ is irrelevant and inflation is effectively viewed as a single-field slow-roll one.
The textbook formulae lead to the expressions for the amplitude $A_\us$ and the spectral index $n_\us$ of the power spectrum of the curvature perturbation:
\beae{
    &A_\us=\frac{1}{24\pi^2\Mpl^4}\frac{V}{\epsilon_\phi}\simeq\frac{\Lambda^4\mu_1^2}{12\pi^2\Mpl^6}, \\
    &n_\us=1-6\epsilon_\phi+2\eta_\phi\simeq1-\frac{4\Mpl^2}{\mu_2^2},
}
which should be $A_\us=\num{2.1e-9}$ and $n_\us=0.965$~\cite{Planck:2018vyg}, namely
\bae{\label{eq: Lambda and mu2}
    \pqty{\frac{\Lambda}{\Mpl}}^4\simeq\num{2.1e-9}\times12\pi^2\pqty{\frac{\mu_1}{\Mpl}}^{-2} \!\!\!\!\!\! \qc 
    \mu_2\simeq10.7\Mpl.
}
The slow-roll parameters $\epsilon_\phi$ and $\eta_\phi$ are defined by
\bae{
    \epsilon_\phi=\frac{\Mpl^2}{2}\pqty{\frac{V_\phi}{V}}^2\simeq\frac{\Mpl^2}{2\mu_1^2} \qc
    \eta_\phi=\Mpl^2\frac{V_{\phi\phi}}{V}\simeq-\frac{2\Mpl^2}{\mu_2^2},
}
and we supposed $\epsilon_\phi\ll\abs{\eta_\phi}$ as is usual in hybrid inflation.
Note that $\phi$'s field value is hence well approximated by the critical value $\phi_\uc$ over the whole relevant dynamics.
Since $\epsilon_\phi$ is very small, the tensor-to-scalar ratio $r$ is also neglgibly small, as calculated by
\bae{
    r\simeq16\epsilon_\phi=\num{8e-14}\pqty{\frac{\mu_1}{10^7\Mpl}}^{-2},
}
for which we used a typical value $\mu_1\sim10^7\Mpl$ adopted in this paper as a reference (see Table~\ref{tab: model parameters}).

The remaining two degrees of freedom further degenerate for the duration of the waterfall phase $N_\water$ and the peak amplitude of the power spectrum there $\calP_\zeta^\text{(peak)}$ as pointed out in Refs.~\cite{Clesse:2015wea,Kawasaki:2015ppx} and improved in TY23a~\cite{Tada:2023pue}.
There, the stochastic noise on the waterfall fields is taken into account only in the valley phase to estimate the initial condition at the critical point, and then the ordinary perturbative approach is applied to the waterfall phase, as a semi-perturbative approach. 
According to their results, both $N_\water$ and $\calP_\zeta^\text{(peak)}$ are mainly controlled by a specific combination of the parameters
\bae{\label{eq: Pi}
    \Pi\coloneqq\frac{M\sqrt{\mu_1\phi_\uc}}{\Mpl^2},
}
as
\bae{\label{eq: Nwater}
    N_\water\simeq\Pi\pqty{\frac{\sqrt{\chi_2}}{2}+\frac{c}{4\sqrt{\chi_2}}}-\Pi^2\frac{\Mpl^2}{12\mu_2^2}(\chi_2+c),
}
and
\bae{\label{eq: calPpeak}
    \calP_\zeta^\text{(peak)}\simeq\frac{\Pi}{2\sqrt{2\pi}\chi_2n}\pqty{1-\frac{\Mpl^2}{3\mu_2^2}\Pi\sqrt{\chi_2}},
}
with a weak logarithmic dependence
\bae{
    \chi_2\coloneqq\ln\frac{M\sqrt{\phi_\uc}}{2\sqrt{n}\sigma_\psi\sqrt{\mu_1}} \qc
    \sigma_\psi\coloneqq\frac{\Lambda^2\sqrt{\Pi}}{4\sqrt{3}(2\pi^3)^{1/4}\Mpl},
}
a fixed parameter $\mu_2\simeq10.7\Mpl$, and a parameter $c\sim1$ representing the effect of the end of inflation.
One finds that $\chi_2\sim10$ for our relevant parameter range $10\lesssim\Pi^2\lesssim1000$.
$\sigma_\psi$ represents the typical amplitude of each waterfall field $\psi_i$ at $\phi_\uc$ due to the stochastic noise.
We will use this value for the initial condition in the stochastic-$\delta\calN$ approach.

A key feature of hybrid inflation is that the peak power spectrum~\eqref{eq: calPpeak} depends on the number of the waterfall fields $n$ while the duration of the waterfall phase~\eqref{eq: Nwater} does not.
Hence, the number of the waterfall fields is expected to affect the statistical properties of the perturbation, keeping the background dynamics intact.
In this work, we run the stochastic-$\delta\calN$ for $n=2$ and $n=15$ with the same values of the other potential parameters as summarised in Table~\ref{tab: model parameters} (dubbed ``Quadratic $n=2$'' and ``Quadratic $n=15$'', respectively).
The initial field values are set to $\phi(t=0)=\phi_\uc$, $\psi_i(t=0)=\sigma_\psi$, and $\dot{\phi}(t=0)=\dot{\psi}_i(t=0)=0$ for $i=1,2,\dots,n$.
We define the end surface by $\eta_\ur=-2$ with
\bae{
    \eta_\ur=\Mpl^2\frac{V_{\psi_\ur\psi_\ur}}{V},
}
for simplicity as done in the literature~\cite{Kodama:2011vs,Clesse:2015wea}, though strictly speaking, the end surface of the $\delta N$ formalism should be a uniform energy density surface~\cite{Lyth:2004gb} (see, e.g., Ref.~\cite{Fujita:2014tja} for a detailed prescription).
As expected, the variance of the first passage time $\expval{\delta\calN^2}=\expval{(\calN-\expval{\calN})^2}$ is significantly suppressed for $n=15$ compared to $n=2$, while the average duration of the waterfall phase $\expval{\calN}$ is almost the same for both cases, as shown in Table~\ref{tab: model parameters}.

\begin{table*}
    \caption{The potential parameter values for the stochastic-$\delta\calN$ computations in the cases of ``Quadratic $n=2$'', ``Quadratic $n=15$'' (see Eq.~\eqref{eq: v quadratic}), and ``Cubic $n=2$'' (see Eq.~\eqref{eq: v cubic}). In all cases, we adopt $M=\phi_\uc/\sqrt{2}=0.01\Mpl$. $\expval{\calN}$ is the average duration of the waterfall phase and $\expval{\delta\calN^2}$ is its variance calculated with $10^7$ samples. We also show the corresponding values of the waterfall e-folds $N_\water$ and the peak power spectrum $\calP_\zeta^\text{(peak)}$ by TY23a's formulae~\eqref{eq: Nwater} and \eqref{eq: calPpeak} for ``Quadratic'' and Eqs.~\eqref{eq: Nwater cubic} and \eqref{eq: calPpeak cubic} for ``Cubic'' for comparison. The end-of-inflation parameter $c$ is set to unity for simplicity.
    }
    \label{tab: model parameters}
    \begin{ruledtabular}
        \begin{tabular}{lccccccccccc}
             & \multicolumn{5}{c}{Parameters} & \quad & \multicolumn{2}{c}{Stochastic-$\delta\calN$} & \quad & \multicolumn{2}{c}{TY23a} \\
            Model & $\Pi$ & $\Lambda/\Mpl$ & $\mu_1/\Mpl$ & $\mu_2/\Mpl$ & $\mu_3/\Mpl$ & & $\expval{\calN}$ & $\expval{\delta\calN^2}$ & & $N_\water$ & $\calP_\zeta^\text{(peak)}$ \\
            \colrule
            Quadratic $n=2$ & $10$ & $\num{2.66e-6}$
            & $\num{7.07e7}$ & $10.7$ & - & & $19.1$ & $0.164$ & & $16.3$ & $0.0845$ \\
            Quadratic $n=15$ & $10$ & $\num{2.66e-6}$
            & $\num{7.07e7}$ & $10.7$ & - & & $18.2$ & $0.0170$ & & $15.6$ & $0.0125$ \\
            Cubic $n=2$ & $11$ & $\num{2.41e-6}$
            & $\num{8.56e7}$ & $6.8$ & $0.0375$ & & $18.4$ & $0.0152$ & & $15.9$ & $0.0702$
        \end{tabular}
    \end{ruledtabular}
\end{table*}

\begin{figure*}
    \centering
    \begin{tabular}{c}
        \begin{minipage}{0.48\hsize}
            \centering
            \includegraphics[width=0.95\hsize]{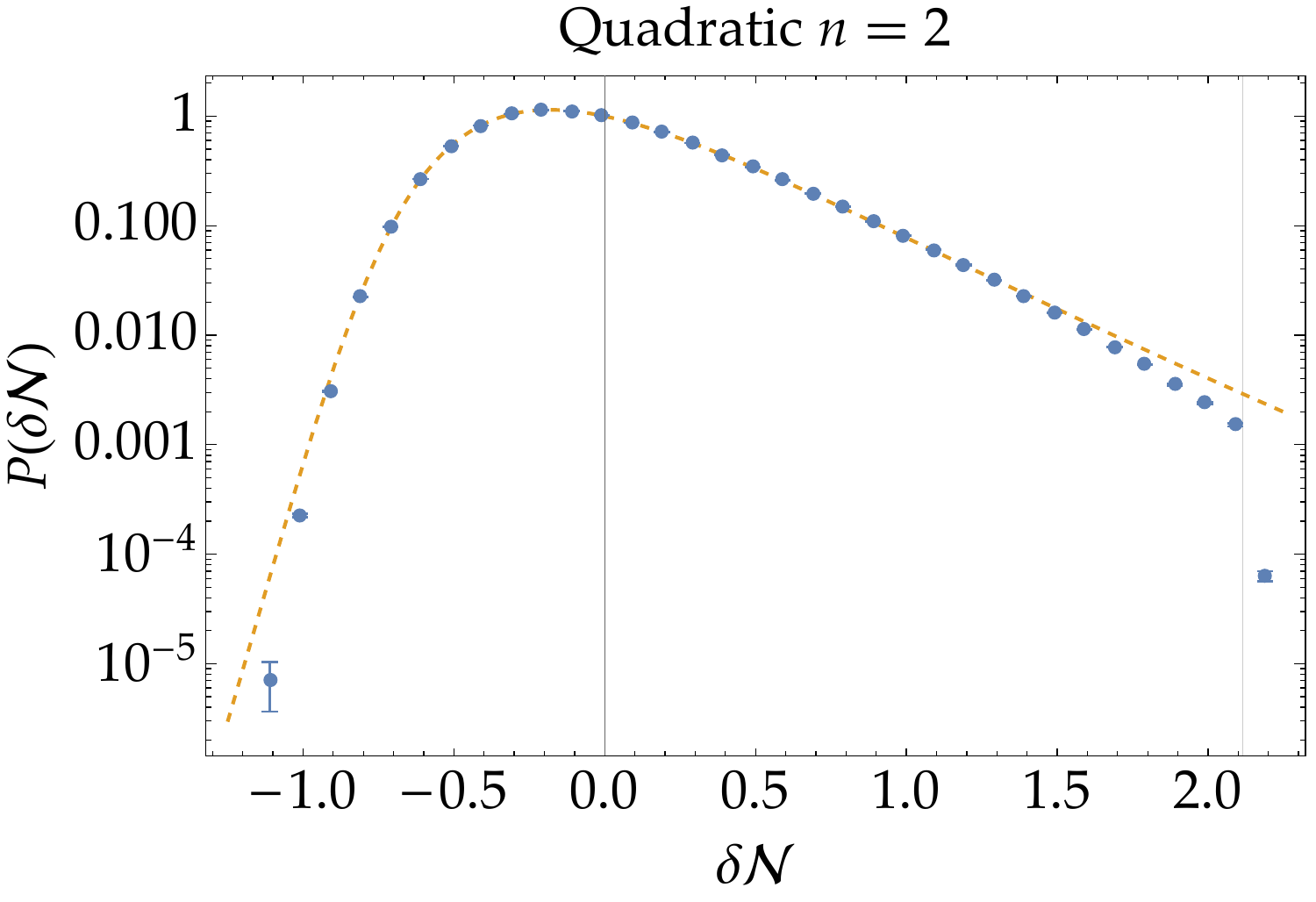}
        \end{minipage}
        \begin{minipage}{0.48\hsize}
            \centering
            \includegraphics[width=0.95\hsize]{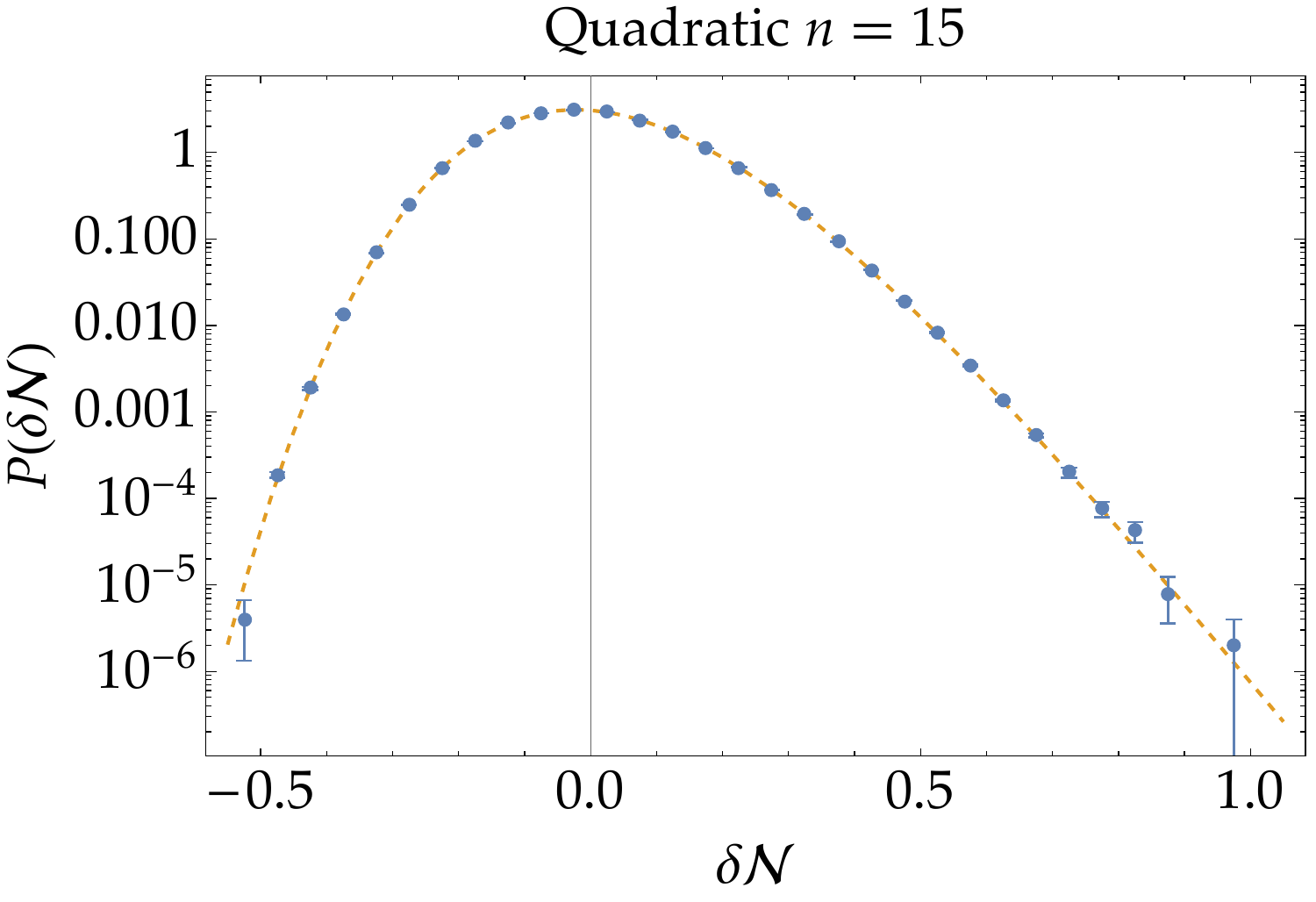}
        \end{minipage} \\
        \begin{minipage}{0.48\hsize}
            \centering
            \includegraphics[width=0.95\hsize]{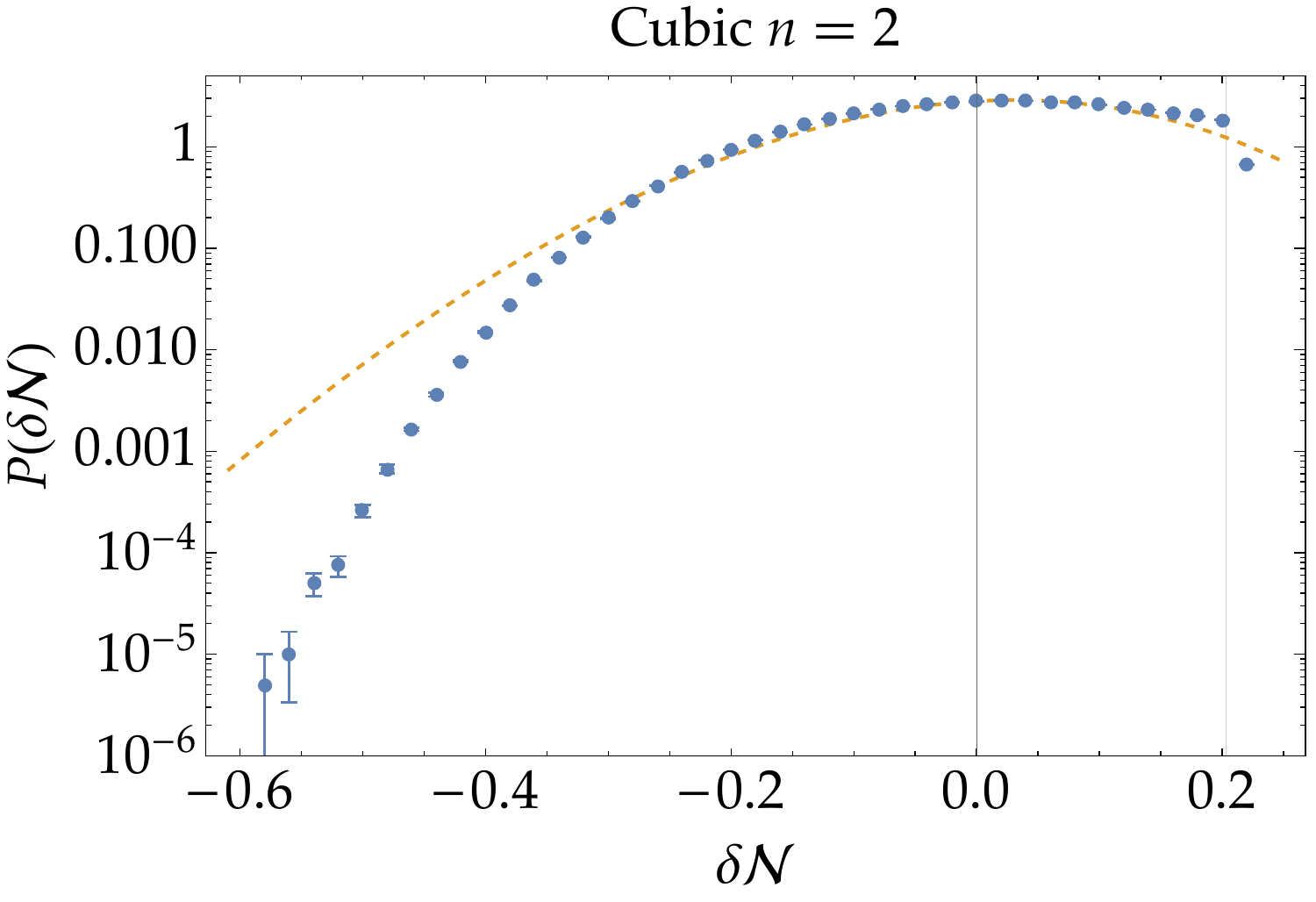}
        \end{minipage}
    \end{tabular}
    \caption{The \acp{PDF} of the curvature perturbation $\zeta=\delta\calN=\calN-\expval{\calN}$ for ``Quadratic $n=2$'' (top-left), ``Quadratic $n=15$'' (top-right), and ``Cubic $n=2$'' (bottom) with the model parameters listed in Table~\ref{tab: model parameters}.
    Blue dots are numerical results with $10^7$ samples. Error bars are estimated by the jackknife resampling. That is, the $10^7$ samples are divided into ten data sets of $10^6$ samples. The \ac{PDF} is computed for each data set and the error is estimated by the standard error for those ten \ac{PDF} data. Orange dashed lines represent Johnson's $S_U$-distribution fitting, whose fitting parameters are listed in Table~\ref{tab: Johnson parameters}.
    The vertical thin lines at $\delta\calN=2.11$ in the top-left panel and at $\delta\calN=0.20$ in the bottom panel correspond to the exact hilltop trajectories $\psi_\ur=0$, which represent effective upper bounds of $\delta\calN$. 
    In fact, the \ac{PDF} value significantly drops beyond these bounds, and the small enough error bars ensure the robustness of these behaviours.
    The $n=15$ case also has a similar upper bound but at $\delta\calN=3.02$ and hence out of the plot range.
    }
    \label{fig: PDF}
\end{figure*}

The top two panels of Fig.~\ref{fig: PDF} show the corresponding \acp{PDF} of the curvature perturbation $\zeta=\delta\calN$ for $n=2$ and $n=15$.
Though both cases exhibit the exponential-tail feature towards the positive and large $\delta\calN$ region, their profiles are distinct from each other.
In particular, the exponent of the tail at $\delta\calN=1$, $\eigenL=-\eval{\dv*{\ln P(\delta\calN)}{\delta\calN}}_{\delta\calN=1}$, is $3.00$ for $n=2$ and $20.9$ for $n=15$.
The fitting parameters with Johnson's $S_U$-distribution~\eqref{eq: Johnson PDF} are summarised in Table~\ref{tab: Johnson parameters}.
Note that the fitting is done in the linear scale, so the parameters are determined almost entirely by the peak behaviour.
It is hence intriguing that the tail behaviour is also well captured by the peak profile in the ``Quadratic'' cases.

\begin{table*}
    \caption{The fitting parameters $\mu$, $\sigma$, $\gamma$, and $\rho$ of Johnson's $S_U$-distribution~\eqref{eq: Johnson PDF} for the \acp{PDF} shown in Fig.~\ref{fig: PDF}. The case ``Cubic $n=2$'' fails to be fit as indicated by the huge errors of the estimated parameters.}
    \label{tab: Johnson parameters}
    \begin{ruledtabular}
        \begin{tabular}{lcccc}
            Model & $\mu$ & $\sigma$ & $\gamma$ & $\rho$ \\
            \hline
            Quadratic $n=2$ & $-0.842\pm0.013$ & $0.454\pm0.006$ & $-3.178\pm0.089$ & $2.429\pm0.025$ \\
            Quadratic $n=15$ & $-0.682\pm0.012$ & $0.530\pm0.003$ & $-7.06\pm0.18$ & $6.65\pm0.06$ \\
            Cubic $n=2$ & $2.4\pm62$ & $0.05\pm1565$ & $78\pm(\num{5.7e5})$ & $17\pm221$
        \end{tabular}
    \end{ruledtabular}
\end{table*}

The $n=2$ case also shows a cutoff at $\delta\calN=2.11$.
It implies that there is an effective maximal e-folding number realised in inflation due to the stochastic noise.
It actually corresponds to the first passage time to the end surface along the exact hilltop $\psi_\ur=0$ shown by the vertical thin line in the figure.
Such a process can happen if the stochastic noise always pushes up the waterfall fields toward the hilltop, and is naturally understood as the realisation that the noise on $\psi$'s maximally extended inflation.
The noise on $\phi$ can go beyond this cutoff in principle, but it is very unlikely. The precise position of the cutoff would also mildly depend on the choice of the end surface.
However, the quantitative cutoff feature should be universal, and it is one of our main findings.
The $n=15$ case also has a similar cutoff, but at $\delta\calN=3.02$ and simply out of the plot range.

\subsection{Including the cubic term}

Another way to resolve the degeneracy is to include the cubic term in the expansion of the inflaton potential as
\bae{\label{eq: v cubic}
    v(\phi)=\frac{\phi-\phi_\uc}{\mu_1}-\frac{(\phi-\phi_\uc)^2}{\mu_2^2}+\frac{(\phi-\phi_\uc)^3}{\mu_3^3}.
}
It changes the spectral index expression as
\bae{
    n_\us\simeq1-\frac{4\Mpl^2}{\mu_2^2}+\frac{12\Mpl^2(\phi_*-\phi_\uc)}{\mu_3^3}.
}
Here, $\phi_*$ corresponds to the pivot scale $k_*=\SI{0.05}{Mpc^{-1}}$ for which we associate $N_*=50\,\text{e-folds}$ before the end surface for simplicity.
This change can make $\mu_2$ free from the constraint value $\mu_2\simeq10.7\Mpl$ and can resolve the degeneracy between $\expval{\calN}$ and $\expval{\delta\calN^2}$, though it requires a mild tuning.
It also means that one can keep the same $\expval{\calN}$ and $\expval{\delta\calN^2}$ while changing the number of the waterfall fields $n$.
In the bottom panel of Fig.~\ref{fig: PDF}, we show the \ac{PDF} of $\delta\calN$ for $n=2$ with tuned parameters shown in Table~\ref{tab: model parameters} (dubbed ``Cubic $n=2$'') so that $\expval{\calN}$ and $\expval{\delta\calN^2}$ are comparable to those in the $n=15$ case up to the quadratic term.
One finds that though $\expval{\calN}$ and $\expval{\delta\calN^2}$ are similar to the $n=15$ case, the \ac{PDF} profile is significantly different.
The upper bound corresponding to the exact hilltop much reduces as $\delta\calN=0.20$ from the ``Quadratic $n=2$" case, and this is the reason for the small variance of the curvature perturbation.
It is a completely different mechanism for small perturbations from the ``Quadratic $n=15$" case, for which the \ac{PDF} is smooth and shows an exponential tail.
Due to the low cutoff, the ``Cubic $n=2$" case with the selected parameters does not expect a sizable amount of \ac{PBH} formation as it requires an order-unity perturbation, while the ``Quadratic $n=15$" does, even though they predict similar average $\expval{\calN}$ and variance $\expval{\delta\calN^2}$.
Furthermore, the ``Cubic $n=2$" case fails to be fit by Johnson's $S_U$-distribution as indicated by large errors for the estimated parameters shown in Table~\ref{tab: Johnson parameters}.

\subsection{Comparison with TY23a}

Though TY23a also predicts the reduction of the perturbation in the ``Cubic'' model, the stochastic-$\delta\calN$ finds much stronger suppression.
TY23a generalised the quadratic calculation to the cubic order as
\bme{\label{eq: Nwater cubic}
    N_\water\simeq\Pi\pqty{\frac{\sqrt{\chi_2}}{2}+\frac{c}{4\sqrt{\chi_2}}}-\Pi^2\frac{\Mpl^2}{12\mu_2^2}(\chi_2+c) \\
    -\Pi^3\frac{\Mpl^4}{32\mu_1\mu_3^3}\pqty{\chi_2^{3/2}+\frac{3c}{2}\chi_2^{1/2}},
}
and
\bae{\label{eq: calPpeak cubic}
    \calP_\zeta^\text{(peak)}\simeq\frac{\Pi}{2\sqrt{2\pi}\chi_2n}\pqty{1-\frac{\Mpl^2}{3\mu_2^2}\Pi\sqrt{\chi_2}-\frac{3\Mpl^4}{16\mu_1\mu_3^3}\Pi^2\chi_2}.
}
In Table~\ref{tab: model parameters}, we show the resultant values of $N_\water$ and $\calP_\zeta^\text{(peak)}$ for given parameters by Eqs.~\eqref{eq: Nwater} and \eqref{eq: calPpeak} for the ``Quadratic'' models and Eqs.~\eqref{eq: Nwater cubic} and \eqref{eq: calPpeak cubic} for the ``Cubic'' model.
While they trace qualitative behaviours of the stochastic-$\delta\calN$ results in the ``Quadratic'' models, TY23a's formulae predict much larger perturbations in the ``Cubic'' model because their semi-perturbative approach fails to capture the upper-bound feature.
It indicates that a full non-perturbative treatment, such as the stochastic approach, is necessary in the ``Cubic'' model.

\section{Discussion and Conclusions}\label{sec: D and C}

In this paper, we investigated the tail behaviour of the \ac{PDF} of the curvature perturbation in the mild-waterfall types of hybrid inflation with the use of the stochastic formalism.
In this formalism, the curvature perturbation is identified with the fluctuation in the e-folding number of inflation, $\delta\calN$, due to the stochastic noise on the inflaton $\phi$ and waterfall fields $\bm{\psi}$.
In the ``Quadratic'' models where the third and higher-order terms in the Taylor expansion of the inflaton potential are irrelevant, the \ac{PDF} shows the exponential tail behaviour as can be seen in the top two panels of Fig.~\ref{fig: PDF}.
On top of this tail, the top-left panel shows a practical upper bound on $\delta\calN$ at $\delta\calN=2.11$ for the model with $n=2$ waterfall fields, which corresponds to the first passage time to the end surface along the exact hilltop $\bm{\psi}=\bm{0}$.
In our ``Cubic'' model, where the third-order term is intentionally set to be as large as the second-order term, the shape of the \ac{PDF} can be significantly altered.
As shown in the bottom panel of Fig.~\ref{fig: PDF}, the upper bound can be lowered to $\delta\calN=0.20$, and then the variance $\expval{\delta\calN^2}$ can be as small as that in the ``Quadratic'' and $n=15$ waterfall-field case.

Let us discuss the implications of these results in comparison with the literature, in particular Refs.~\cite{Kawasaki:2015ppx,Tada:2023fvd,Tada:2023pue,Tada:2024ckk,Moursy:2025ljr}.
First of all, it is shown that the ``Quadratic'' models have degeneracy in the parameter degrees of freedom for the average $\expval{\calN}$ and the variance $\expval{\delta\calN^2}$ of the duration of the waterfall phase, so that one is fixed if the other is fixed~\cite{Clesse:2015wea,Kawasaki:2015ppx}.
From the perspective of consequent astrophysical objects, the average $\expval{\calN}$ corresponds to the mass of the collapsed objects while the variance $\expval{\delta\calN^2}$ controls their abundance.
Namely, the mass function of the collapsed objects is (almost) uniquely determined in the ``Quadratic'' models irrespective of the parameter choices.
Ref.~\cite{Kawasaki:2015ppx} has calculated the momenta of $\delta\calN$ up to the kurtosis, and concluded that massive enough \acp{PBH} are unavoidably overproduced in the ``Quadratic'' models with a few waterfall fields.
Since the exponential tail enhances the \ac{PBH} abundance, our results strengthen this conclusion.
On the other hand, the amplitude of the so-called \acp{SIGW} is almost determined by the peak behaviour of the \ac{PDF}, i.e., the variance, and hence the conclusion of Refs.~\cite{Tada:2024ckk,Moursy:2025ljr} about the \ac{SIGW} in GUT-Higgs hybrid inflation would not be affected so much.

One way to resolve the degeneracy is to increase the number of the waterfall fields~\cite{Halpern:2014mca}.
The power spectrum (or the variance) of the curvature perturbation is expected to be (almost) inversely proportional to the number of the waterfall fields $n$ as can be seen in Eq.~\eqref{eq: calPpeak}, keeping the average $\expval{\calN}$ almost unchanged.
Hence, the large number of the waterfall fields can avoid the overproduction of massive \acp{PBH}.
Calculating the power spectrum in the stochastic-$\delta\calN$ approach, Ref.~\cite{Tada:2023fvd} concluded that $n\sim5$ waterfall fields can lead to an appropriate amount of \acp{PBH} produced.
Though the exponential tail would increase the required number of the waterfall fields, there should be a certain number of $n$ for appropriate \ac{PBH} production.
The detailed amount of \acp{PBH} produced depends not only on the \ac{PDF} but also on the spatial profile of overdensities~\cite{Atal:2019erb,Escriva:2019phb}, and hence it requires, e.g., lattice simulations of inflation~\cite{Caravano:2021pgc,Mizuguchi:2024kbl,Caravano:2025klk,STOLAS} which we leave for future work.

Another way to resolve the degeneracy is to include the cubic term in the inflaton potential.
TY23a~\cite{Tada:2023pue} showed in a semi-perturbative way that a fine-tuned choice of the cubic term can reduce the amplitude of the power spectrum. 
It was then concluded that \ac{PBH} production with an appropriate abundance is possible even with a small number of the waterfall fields.
However, the bottom panel of Fig.~\ref{fig: PDF} indicates that the upper bound on the curvature perturbation is significantly lowered in this case, and the perturbation variance becomes much smaller than TY23a predicted, as shown in Table~\ref{tab: model parameters}.
Furthermore, as the upper bound is smaller than unity, the realisation probability of the order-unity curvature perturbation relevant for \ac{PBH} formation is much smaller than the expected value either in the Gaussian distribution or in the exponential-tail one.
Hence, we conclude that \acp{PBH} almost never form in this parameter set, contrary to the expectation in TY23a.
The production of a sizable amount of \acp{PBH} may be possible if the upper bound lies just around the \ac{PBH} threshold $\delta\calN\sim1$, but it requires extremely fine tuning of the parameters and unrealistically high precision of numerical computations.

Such a low upper bound of the curvature perturbation itself may lead to an interesting astrophysical consequence in, e.g., the halo formation apart from \acp{PBH}.
To our best knowledge, the formation of collapsed objects from a bounded initial perturbation has not been studied well.
Such a bound is expected to change the spatial profile of overdensities.
It also simply changes the balance between the over- and underdensities. 
The \ac{PDF} shown in the bottom panel of Fig.~\ref{fig: PDF} corresponds to the significant deviation of the overdensity (or underdensity) probability from $50\%$:
\bae{
    \int_0^\infty P(\delta\calN)\dd{\delta\calN}\simeq0.533>0.5.
}
The $N$-body simulations with such a bounded perturbation may be an interesting direction of future work.

We also emphasise the usefulness of Johnson's $S_U$-distribution we introduced in Sec.~\ref{sec: SU}.
As can be seen in the top two panels of Fig.~\ref{fig: PDF}, it can well fit both the peak and the tail behaviours of the \ac{PDF}, and is expected to work for a wide class of heavy tails.
It also reveals the mapping between the Gaussian field and the curvature perturbation as given in Eq.~\eqref{eq: g to dN}.
Such a mapping is helpful to apply the peak theory (see Refs.~\cite{Yoo:2019pma,Kitajima:2021fpq,Inui:2024fgk}).
As it does not include the information of the spatial distribution, it should be calculated separately, e.g., in lattice simulations.

\acknowledgments
T.M. is supported by Specific Project Grant from TMCIT.
Y.T. is supported by JSPS KAKENHI Grant
No.~JP24K07047.


\bibliography{main}
\end{document}